# From Hubs to Deserts: Urban Cultural Accessibility Patterns with Explainable AI

### Protik Bose Pranto
Ira A. Fulton Schools of Engineering
Arizona State University
Tempe, AZ, USA
ppranto@asu.edu

### Minhazul Islam
Ira A. Fulton Schools of Engineering
Arizona State University
Tempe, AZ, USA
mislam23@asu.edu

### Ripon Kumar Saha
Ira A. Fulton Schools of Engineering
Arizona State University
Tempe, AZ, USA
rsaha8@asu.edu

### Abimelec Mercado Rivera
The Polytechnic School
Arizona State University
Mesa, AZ, USA
abimelec.mercado@asu.edu

### Namig Abbasov
Data Science and Analytics Unit
Arizona State University
Tempe, AZ, USA
nabbaso1@asu.edu

## Abstract

Cultural infrastructures, such as libraries, museums, theaters, and galleries, support learning, civic life, health, and local economies, yet access is uneven across cities. We present a novel, scalable, and open-data framework to measure spatial equity in cultural access. We map cultural infrastructures and compute a metric called Cultural Infrastructure Accessibility Score (CIAS) using exponential distance decay at fine spatial resolution, then aggregate the score per capita and integrate socio-demographic indicators. Interpretable tree-ensemble models with SHapley Additive exPlanation (SHAP) are used to explain associations between accessibility, income, density, and tract-level racial/ethnic composition. Results show a pronounced core–periphery gradient, where non-library cultural infrastructures cluster near urban cores, while libraries track density and provide broader coverage. Non-library accessibility is modestly higher in higher-income tracts, and library accessibility is slightly higher in denser, lower-income areas.

## 1 Introduction

Cultural infrastructure is integral to equitable, inclusive, and vibrant urban environments. Beyond artistic expression and heritage preservation, assets such as libraries, museums, theaters, and art galleries support educational attainment, civic participation, mental health, and local economic development [10, 37]. Framed as the "fourth pillar" of sustainable development alongside economic, social, and environmental goals [4], culture can foster inclusive growth and social cohesion [34].

Equitable access is simultaneously about cultural vitality and social fairness. All residents should be able to benefit from the educational, civic, and health advantages of cultural engagement. When neighborhoods lack cultural resources, these benefits are foreclosed, reinforcing social and spatial inequalities. Yet cities around the world exhibit uneven distributions of these infrastructures. Many marginalized neighborhoods are frequently excluded from cultural investment and planning [21, 36], creating "cultural deserts" - regions with limited access to cultural amenities. On the other hand, affluent or central districts are more concentrated with cultural infrastructure relative to other regions. This spatial imbalance mirrors and reinforces broader urban disparities and warrants urgent attention in research and planning. Despite broad recognition, planning and policy still struggle to put equal cultural accessibility into practice. Many initiatives prioritize the economic returns of cultural investment (creative economy, tourism) over the basic question of whether residents can reach and use facilities [20]. Empirically, few analyses identify where cultural infrastructure is over- or underserved relative to the local population and need. Where measures do exist, they often reduce access to venue counts or simple proximity/density, ignore socio-demographic context, and rely on ad-hoc methods that limit comparison across places over time [11, 29, 30]. Since most new cultural venues are sited to maximize revenue and visibility, they cluster where demand, transit access, and philanthropic support are strongest (the urban core). Outlying neighborhoods, with weaker expected returns, receive fewer investments, which amplifies center–periphery disparities. These shortcomings motivate a citywide, open-data, multi-asset framework that quantifies accessibility, incorporates demographic context, and yields reproducible, interpretable equity metrics.

To address these challenges, we introduce a citywide, scalable framework that integrates open urban data, socio-demographic indicators, spatial analysis, and interpretable machine learning to evaluate cultural accessibility. The workflow proceeds in four steps. First, we map cultural infrastructure by compiling geospatial records of libraries, museums, theaters, and galleries from public databases and open data portals to build a citywide inventory. Second, we integrate community context by joining socio-economic and demographic indicators (e.g., household income, racial/ethnic composition, educational attainment, population density, with optional health or transit metrics) from census and other open sources at a common spatial unit (tract, neighborhood, or grid). Third, we compute accessibility by estimating a Cultural Infrastructure Accessibility Score (CIAS) with an exponential distance–decay function (giving greater weight to nearer sites), aggregating to a common unit, and deriving a per-capita Socio-Cultural Accessibility Score (SCAS) to flag over- and underserved areas and reveal clusters and gaps. Finally, we interpret patterns with an interpretable model to relate accessibility to neighborhood context, clarifying factors associated with inequitable access.

We center the analysis on median household income. Although many outcomes could be examined, income offers a direct lens on the well-documented clustering of cultural venues in affluent cores. Accordingly, we pose two research questions to guide our study of cultural infrastructure equity and its socio-spatial correlates:

(1) **Identification of disparities:** Which neighborhoods and racial/ethnic subgroups are over-served, and which are under-served, in their access to cultural infrastructure (as measured by the socio-cultural accessibility score)?

(2) **Determinants and differential effects:** Controlling for spatial structure, how do socio-demographic composition (race or ethnicity shares, population density) and socio-cultural accessibility relate to household income, and do these associations differ across cultural asset types and ethnic groups?

We validate the framework in New York City (NYC), selected for its diverse socio-economic context, income, and cultural assets, to address our two research questions. The results yield a coherent equity narrative: cultural access in NYC follows a core–periphery pattern with clear heterogeneity by asset type and population group. Median household income aligns mostly with demographic composition and population density. Proximity to cultural assets shows a smaller, asset-specific association. The framework is designed to be adaptable to any metropolitan area with access to open geographic and census-style datasets. This offers an actionable basis for equity-focused cultural planning so cultural vitality becomes a shared asset rather than a privilege. By answering these questions, we clarify the landscape of cultural access inequity and highlight the factors most closely associated with whether communities enjoy rich cultural opportunities or are left behind. These insights are intended to inform more equitable cultural policy and contribute to broader debates on inclusive and sustainable urban development.

## 2 Literature Review

Researchers measure cultural access using standard spatial tools: proximity to the nearest venue, density or coverage within buffers or catchments [20], and travel-time accessibility (e.g., 15-minute walk/ride, two-step floating catchment area) [15, 17]. Many studies combine these into composite indices that also include transit availability and affordability. Such approaches help identify over-served areas (where supply is high relative to population) and under-served areas (with few nearby options), often termed as cultural "deserts" or "cold spots." Traditional quantitative models relate museum attendance to personal characteristics and accessibility indices [2]. This enables researchers to isolate the effect of physical access on cultural participation while controlling for variables such as education, class, and ethnicity. Other studies have used spatial regression techniques (e.g., spatial lag) to account for geographic autocorrelation when analyzing the distribution of cultural facilities [12].

Recent years have seen the emergence of big geospatial datasets, enabling more data-driven and algorithmic analyses of cultural access. Researchers apply point-of-interest (POI) data mining and clustering algorithms to detect spatial patterns. For example, one study gathered extensive POIs for libraries, art galleries, and museums via web APIs, applying DBSCAN clustering and kernel density analysis to identify cultural clusters and voids [12]. Similarly, Ma et al. (2025) used the GeoDetector tool to quantify factor effects and interactions on infrastructure distribution [18]. This is similar to feature importance analysis in machine learning, but grounded in spatial statistics. Innovative techniques are also being developed to integrate qualitative dimensions into accessibility models. A study of library accessibility created a "Library Attraction Index" using topic modeling and sentiment analysis of social media data. This index weighted libraries by their perceived quality and appeal, which was then incorporated into an enhanced two-step floating catchment area model to measure effective accessibility [16]. This illustrates a broader trend: ML and AI augment GIS by processing mobility data, interpreting user reviews, and improving allocation.

The existing literature shows strong spatial disparities in cultural infrastructure tied to socioeconomic and racial inequalities. Cultural venues cluster in affluent, central urban areas, leaving peripheral and marginalized neighborhoods with fewer options and lower participation [14]. While prior research has effectively used GIS and statistical methods to reveal these patterns and advocate for more equitable cultural planning, several limitations persist. Many studies focus on one city or one asset type and struggle to disentangle the complex interplay of contributing factors. We address these gaps by integrating multiple asset classes into a composite distance-decay metric, defining over- and under-service with clear distributional thresholds, computing race/ethnicity-specific per-capita accessibility, and pairing spatial metrics with explainable AI. This helps separate the roles of economics, urban form, and cultural proximity, and their directional effects, providing a stronger basis for targeted policy interventions.

## 3 Methodology

### 3.1 Spatial Mapping and Area Refinement

*3.1.1* **Geographic Grid Creation:** The framework applies a high-resolution spatial sampling strategy by constructing a uniform grid across the study area [9, 26, 32]. The grid consists of evenly spaced latitude-longitude points based on a predefined geographic bounding box and resolution. The grid points are calculated by dividing the latitude and longitude ranges of the study area into evenly spaced intervals, with each grid cell representing a fixed spatial unit. The resolution of the grid determines the number of points in the grid, with higher resolutions offering finer spatial granularity. The choice of resolution depends on the size of the study area and the level of detail required for the analysis. This method enables consistent and standardized comparisons across the area. This grid-based approach ensures comprehensive spatial coverage and mitigates potential biases associated with administrative boundaries, such as irregular political or census divisions [19, 22]. It captures detailed variations in cultural accessibility and demographic conditions, enabling in-depth spatial analysis and revealing patterns of accessibility with high precision.

*3.1.2 Spatial Filtering:* The framework begins with the spatial filtering of the grid by removing any points that fall outside the defined study area. This filtering step is implemented using GeoJSON-based polygon containment queries [38], which retain only those grid points located within the geographic boundary of the study area. Specifically, the longitude and latitude of each grid point are checked to determine if it lies within the study area's boundary, ensuring that only relevant data are included.

## 3.2 Dataset Preparation

*3.2.1 Cultural Infrastructure Data:* The framework collects cultural infrastructures data, including the locations and types of cultural institutions (e.g., libraries, museums, etc.), from publicly available sources such as city open data portals and OpenStreetMap (OSM) [1, 5]. Each cultural asset is mapped with geographic coordinates (longitude, latitude). This data forms the basis for calculating the cultural accessibility Score, explained in the next subsection.

*3.2.2 Cultural Accessibility Score Computation:* After collecting the cultural infrastructure data, the Cultural Infrastructure Accessibility Score (CIAS) is computed for each grid point. By using an exponential decay function, the CIAS quantifies cultural accessibility by accounting for both the distance to and the relative importance of nearby cultural infrastructures. For each grid point $p_i(x_i, y_i)$, the CIAS, $S(p_i)$, is computed using the following formula:

$$S(p_i) = \sum_{j=1}^{N} w_j e^{-\alpha d_{ij}} \qquad (1)$$

where:

- $i$ corresponds to the index of each grid point
- $j$ corresponds to each cultural infrastructure
- $N$ represents the total number of cultural infrastructures considered in the analysis. This number can vary depending on the analysis; for example, if the analysis focuses only on libraries, $N$ will only represent the number of libraries in the area, rather than all types of cultural sites.
- $d_{ij}$ is the Euclidean distance between grid point $p_i$ and cultural infrastructure $j$,
- $w_j$ is a non-negative weight assigned to cultural infrastructure $j$ (determined based on its relative importance or type),
- $\alpha$ is a decay parameter that controls how rapidly the influence of cultural sites diminishes with increasing distance.

Equation 1 ensures that cultural assets located closer to a grid point have a higher influence on the CIAS than those farther away. The decay parameter $\alpha$ controls how distance affects the accessibility score, adjusting the weight given to more distant cultural assets. To efficiently compute accessibility scores across large datasets, the framework uses KDTree data structures [33], which facilitate fast proximity queries and enhance scalability for extensive datasets. This approach allows quick retrieval of the nearest cultural sites, supporting score calculation even in large metropolitan areas. The CIAS is computed for all cultural sites, resulting in a composite cultural accessibility index for each grid point (Subsection 3.3).

*3.2.3 Data Integration and Feature Assembly:* After computing CIAS scores for each grid point, the next step is to integrate these cultural accessibility scores with demographic, health, and spatial data. This process involves combining multiple datasets to create a unified representation of the study area's attributes.

*Demographic Data:* Demographic data is sourced from publicly available census datasets [8], aggregated at the census tract level. Key variables that are considered for this study include:

- **Population metrics:** total population, population density,
- **Racial/ethnic composition:** total and percentage population shares of all ethnic groups,
- **Socioeconomic indicators:** median household income.

The framework handles all missing values by imputing them using column-wise means [27] to ensure consistency.

*Public Health Data:* Public health indicators, such as mental health data (e.g., depression prevalence), are sourced from open-access public health datasets [6], and aggregated at the tract level.

*CIAS Score Aggregation:* CIAS is initially computed at a fine spatial scale, at each grid point, reflecting the proximity and influence of cultural infrastructure. However, since other demographic and public health data are typically available at the coarser spatial scale of census tracts, the CIAS scores must be aggregated to this level to ensure consistent spatial alignment across datasets. This aggregation facilitates the integration of cultural accessibility data with socio-demographic and health-related attributes, enabling a comprehensive analysis at the same spatial resolution. To handle edge cases where grid points fall on the boundaries of multiple regions or near multiple administrative units, centroid-based containment checks are employed [7]. This ensures that each grid point is accurately assigned to its nearest administrative region, even when located on or near polygon boundaries. By comparing the centroids of both grid points and census tracts, this approach minimizes the risk of misassignment or exclusion of boundary-adjacent points. Consequently, the CIAS scores are correctly aggregated at the same level, ensuring consistent spatial analysis across the study area.

*Feature Consolidation and Normalization:* Once the data from various sources are spatially aligned, the framework consolidates them into a single unified dataset. To ensure comparability across features, z-score normalization is applied to all continuous variables [3]. The final merged dataset provides a comprehensive set of attributes for each census tract, including cultural infrastructure data (e.g., CIAS for cultural assets), demographic data (e.g., population, income, and racial composition), health data (e.g., mental health indicators), and geospatial data (e.g., geographic coordinates). All grid points (latitude and longitude) within each census tract will share the same demographic, health, and aggregated CIAS data, ensuring consistency across the analysis. This unified dataset allows for a multi-dimensional analysis of accessibility patterns, incorporating socio-cultural, health, and demographic factors, thus enabling detailed spatial analysis of equity across the study area.

## 3.3 Socio-Cultural Accessibility Score (SCAS)

Once the final dataset has been integrated, the next step is the calculation of the **Socio-Cultural Accessibility Score (SCAS)**. The SCAS measures the availability and distribution of cultural assets across different regions, taking into account both the proximity and concentration of cultural infrastructure and the population size in each area. The score is computed at the census tract level to maintain consistency with the demographic and public health data, which are aligned to this spatial scale. SCAS for a given census tract $t$ is defined as the ratio of the total cultural accessibility score within the tract to the total population of the tract, as represented by the following formula:

$$\text{SCAS}_t = \frac{\sum_{j=1}^{N} \text{CIAS}_{tj}}{P_t} \tag{2}$$

Where:

- $\text{SCAS}_t$ is the Socio-Cultural Accessibility Score for the census tract level at index $t$,
- $\text{CIAS}_{tj}$ is the aggregated Cultural Infrastructure Accessibility Score for each cultural site $j$ within census tract $t$,
- $N$ is the total number of cultural sites in census tract $t$,
- $P_t$ is the total population of census tract $t$.

Equation 2 calculates cultural accessibility per capita, providing insights into the relative availability of cultural resources in relation to the population in each census tract. For interpretation, the framework suggests a classification module that partitions tract-level SCAS into equal-frequency quantile bins. The lower, middle, and upper portions of this distribution are referred to as low, moderate, and high access, which allows for the identification of regions that may be underserved or overserved in terms of cultural infrastructure (shown in Figure 1). These thresholds are city-specific and relative, supporting within-city comparisons.

To further refine the analysis, the SCAS can be broken down by specific population subgroups based on racial composition. This enables a more granular understanding of cultural accessibility disparities across different demographic groups. For example, the SCAS formula can be adjusted to focus on a particular racial or ethnic group $r$ within a census tract, as shown below:

$$\text{SCAS}_{t,r} = \frac{\sum_{j=1}^{N} \text{CIAS}_{tj}}{P_{t,r}} \tag{3}$$

Where:

- $\text{SCAS}_{t,r}$ represents the Socio-Cultural Accessibility Score for census tract $t$ and racial group $r$,
- $P_{t,r}$ is the population of racial group $r$ within census tract $t$.

This modification allows for the identification of areas where specific racial or ethnic groups may have disproportionately low or high access to cultural infrastructure. By applying this approach, cultural infrastructure can be more effectively distributed to reduce disparities and improve cultural access for underserved groups. SCAS provides a comprehensive measure of cultural accessibility, incorporating both the physical availability of cultural sites and the demographic characteristics of the population they serve. This score facilitates a detailed analysis of spatial equity and cultural access across the study area, offering valuable insights for policy-making aimed at enhancing socio-cultural equity and access to cultural resources for all population groups.

## 3.4 Predictive Modeling and Analysis

This part outlines the process of using machine learning (ML) models to explore the relationships between various socio-demographic factors, cultural infrastructure accessibility, and public health indicators with median household income.

### 3.4.1 Model Development and Evaluation:
A variety of ML models, including traditional models (e.g., Linear Regression), tree-based models (e.g., Random Forest, Gradient Boosting), and neural networks [28, 31, 35], are used to predict median household income. These models are trained on demographic, cultural, and health data to identify the factors that contribute to variations in income levels. A training and testing split is employed to ensure that the models generalize well to unseen data, mitigating overfitting and improving the robustness of predictions. After training, model performance is evaluated using standard metrics: R-squared ($R^2$), Mean Squared Error (MSE), and Mean Absolute Error (MAE). These metrics help assess the predictive accuracy of the models and enable comparisons to determine the best-performing algorithm. Hyperparameter tuning is applied to optimize model performance, where grid search or other optimization methods are used to select the best configuration of parameters. The evaluation results inform the selection of the most suitable model for subsequent interpretability analysis.

### 3.4.2 Model Interpretability:
For a deeper understanding of how the socio-demographic, cultural, and health-related factors impact household income predictions, interpretability methods such as SHAP (Shapley Additive Explanations [23]) are employed. SHAP values provide insight into the contribution of each feature to the model's predictions, offering transparency and explaining how the model arrives at its decisions. A positive SHAP value means a feature pushes the predicted household income up (above the model baseline), while a negative value pushes it down. By analyzing the feature importance through SHAP, it is possible to gain valuable insights into which factors have the greatest impact on income.

### 3.4.3 Policy Implications:
By combining a machine learning model with SHAP for interpretability, the framework provides a comprehensive analysis of the relationships between cultural accessibility, socio-demographic factors, and public health outcomes with income. The insights gained from this analysis can inform policy decisions aimed at improving access to cultural resources, reducing disparities, and promoting equal cultural accessibility across different population groups and regions in a study area.

## 4 Case Study

We implemented and evaluated our framework in New York City (NYC), USA, chosen for its diversity in population, socio-economic status, cultural infrastructure, and public health indicators. As one of the largest cities globally, NYC is home to over 8 million residents from various ethnic, income, and cultural backgrounds, making it an ideal urban setting for studying disparities in cultural access [8]. In NYC, a borough or county is one of the city's five primary administrative divisions. The five boroughs are **Manhattan** (New York County), **Brooklyn** (Kings County), **Queens** (Queens County), **Bronx** (Bronx County), and **Staten Island** (Richmond County), shown in Figure 1a. We conducted all analyses across these five boroughs (counties). For cultural infrastructure data, we relied on the NYC Open Data Portal, selecting libraries, museums, art galleries, and theaters as cultural infrastructures for our analysis [25]. Demographic and socio-economic data were sourced from the U.S. Census Bureau, while public health indicators (depression rates) were obtained from the NYC Community Mental Health Data [24].

To ensure comprehensive coverage of NYC, we implemented a spatial sampling approach using a 500x500 grid system. All data sources were aggregated at the census tract level to ensure spatial consistency. After constructing the uniform grid, clipping points to the study boundary, and joining cultural, demographic, and health layers, the unified geospatial dataset comprises 74484 grid-point observations with 16 features. Each record includes geographic coordinates (latitude and longitude); four component Cultural Infrastructure Accessibility Scores (CIAS) for art galleries, theaters, libraries, and museums; tract-level total population and population density; a public-health indicator (depression prevalence, in percentage); and tract-level population counts by race following U.S. OMB [13] (Office of Management and Budget) categories (with city-wide percentage of total population): (1) **White** (37.75%), (2) **Black or African American** (23.05%), (3) **American Indian or Alaska Native** (0.61%), referred as "Native" in the paper, (4) **Asian** (14.39%), (5) **Native Hawaiian and Other Pacific Islander** (0.06%), referred as "nhpi" in the paper, (6) **Other single race** (15.26%), and (7) **Two or more races (Multiracial)** (8.88%). These ethnicity based population values are counts (not percentages) and are propagated to all grid points within the containing census tract. We also calculated the Socio-Cultural Accessibility Score (SCAS) to assess cultural accessibility for the entire population and ethnic groups in NYC based on census data. We treated all cultural sites as equally important (setting weight as 1) and used a decay parameter of 0.0005. To identify under- and over-served areas from SCAS, we partition the tract-level SCAS distribution into 20 equal-frequency bins.

Before training the ML models, we computed the Pearson correlations between all variables with racial/ethnic composition measured as tract-level percentages. The dataset was split into a training set (80%) and a testing set (20%) to evaluate model performance. We implemented and compared multiple models, including traditional models (Linear Regression), tree-based models (Random Forest, Gradient Boosting, XGBoost), and neural networks (Fully Connected Neural Network or FCNN). We used grid search and cross-validation techniques to identify the optimal settings for each model. Each model was evaluated based on R-squared ($\mathbb{R}^2$), Mean Squared Error (MSE), and Mean Absolute Error (MAE), to identify the best-performing model for predicting household income. From the SHAP model interpretability, we identified the most influential features and assessed their impact on income prediction.

## 5 Result

### 5.1 Cultural Infrastructure Mapping

*City-wide mapping:* Cultural infrastructures are most densely concentrated in Manhattan, with a pronounced cluster of theaters and museums (Figure 1b). Brooklyn shows a secondary concentration, especially toward the borough's more central areas, while Queens and the Bronx display a more dispersed pattern with fewer high-density clusters. Staten Island has comparatively sparse coverage across all categories. Galleries are numerous and spread farther from the core than theaters and museums, which remain tightly clustered; libraries are the only asset type with near city-wide coverage, extending into peripheral parts of all boroughs.

*Based on Socio-Cultural Accessibility Score (SCAS):* As shown in Figure 1c, considering the total population, the score is highest across much of Manhattan. Brooklyn is mixed, with higher values in several parts of the borough and lower values toward its periphery. Queens is heterogeneous, showing moderate values in some areas and lower values in more outlying sections. The Bronx is largely mid-range with scattered higher and lower pockets, while Staten Island is predominantly low with only limited higher areas.

By population group, the overall gradient persists but with notable differences. For the White population (Figure 1d), values are highest across many tracts in Manhattan, mixed (mostly high) in Brooklyn, moderate and lower toward the edges of Queens, mostly mid-to-low in the Bronx, and predominantly low in Staten Island; the score generally declines with distance from the Manhattan–Brooklyn core. For the Black population (Figure 1e), Manhattan includes many high-scoring tracts, Brooklyn is mixed around the middle of the range, Bronx is largely mid-range, Queens varies but trends lower in more peripheral parts, and Staten Island is mostly low to moderate. For the Asian population (Figure 1f), Manhattan again shows many high values; Brooklyn is mixed with numerous mid-to-high tracts; Queens is highly variable but leans lower across large portions of the county; Bronx is mostly mid-range; and Staten Island is low to moderate (in some tracts). For the "Other" race population (Figure 1g), Manhattan records many of the highest values, Brooklyn is mixed with mid-range tracts and scattered higher pockets, Queens is heterogeneous, Bronx is mostly lower to mid-range, and Staten Island remains low to moderate. For the Multiracial population (Figure 1j), high values are concentrated in Manhattan. Parts of Brooklyn show numerous mid-to-high tracts. Queens is mixed, with pockets of higher scores nearer the core and lower values toward the periphery. The Bronx is largely mid-range with scattered higher pockets. Staten Island is predominantly low. For American Indian/Alaska Native (Figure 1h), and for Native Hawaiian and Other Pacific Islander populations (Figure 1i), the score is visible in relatively few tracts citywide, with the highest values concentrated in parts of Manhattan and scattered in Brooklyn and Queens; Bronx and Staten Island show mostly low to moderate.

### 5.2 Spatial Distribution of Household Income

As shown in Figure 2, Manhattan registers the highest levels overall, with large contiguous stretches in the top range. Brooklyn is heterogeneous: higher-income tracts cluster in the northwest/central portions, while broader belts across the east and south are lower to mid-range. Queens skews mid-range, punctuated by scattered higher-income pockets and more lower–mid tracts toward the outer parts of the county. The Bronx shows the lowest incomes overall, with only limited mid-range areas. Staten Island is predominantly middle-income, showing variation across tracts but fewer extremes than Manhattan or the Bronx. Overall, the map suggests a core-to-edge gradient, with peak incomes in Manhattan and progressively lower values across much of Brooklyn, Queens, and especially the Bronx, while Staten Island sits largely in the middle.

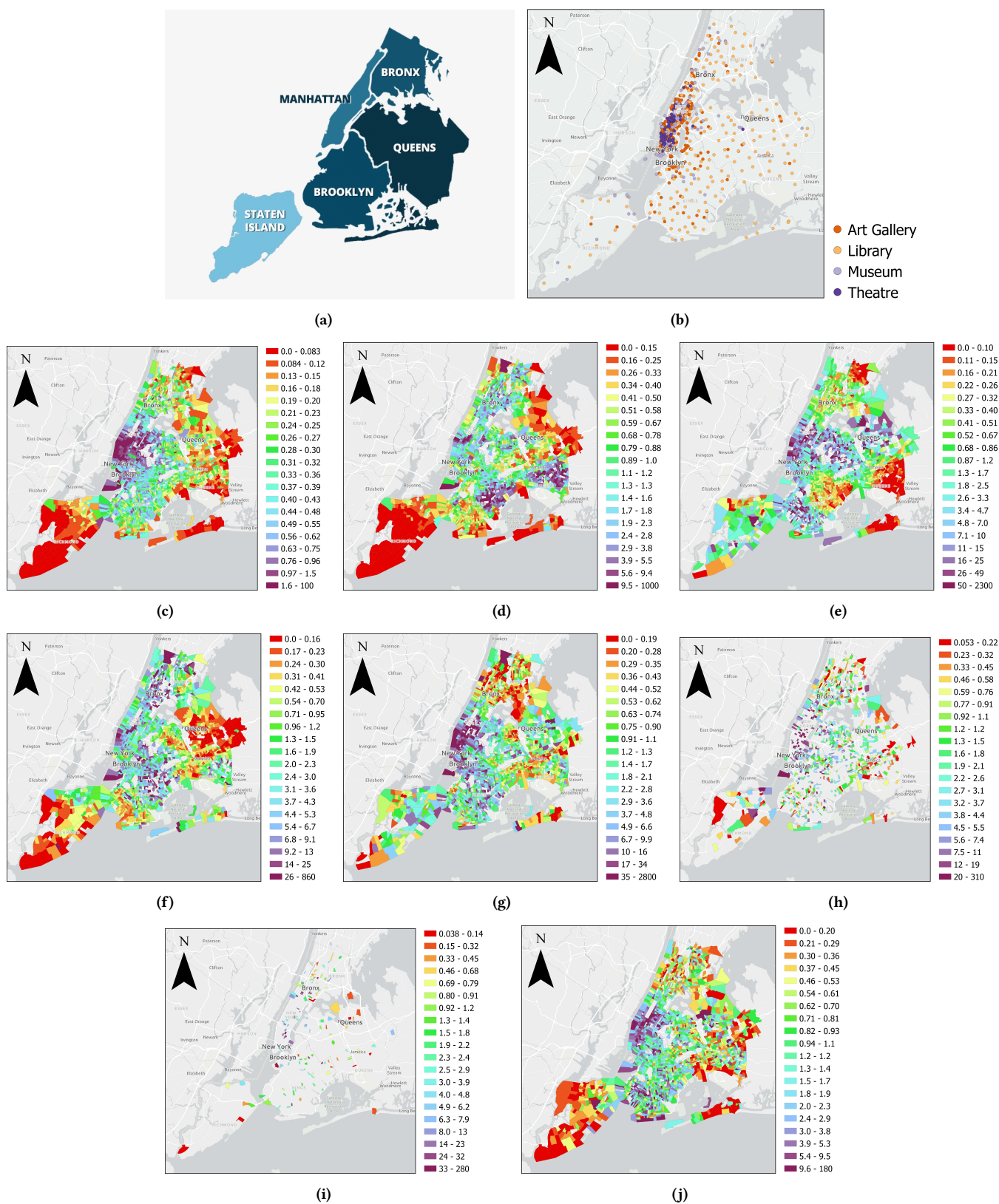

**Figure 1: Geographies and Socio-Cultural Accessibility Score (SCAS) across New York City. (a )County boundaries, (b) locations of cultural infrastructures, (c) total population, (d) White population, (e) Black population, (f) Asian population, (g) Others population, (h) American Indian/Alaska Native population, and (i) Native Hawaiian and Other Pacific Islander population population (j) Multiracial (Two or More Race). All panels use a common color scale. The color bar indicates SCAS, with higher values denoting greater per-capita proximity to cultural infrastructures. Non-color tracts indicate zero population for the corresponding subgroup.**

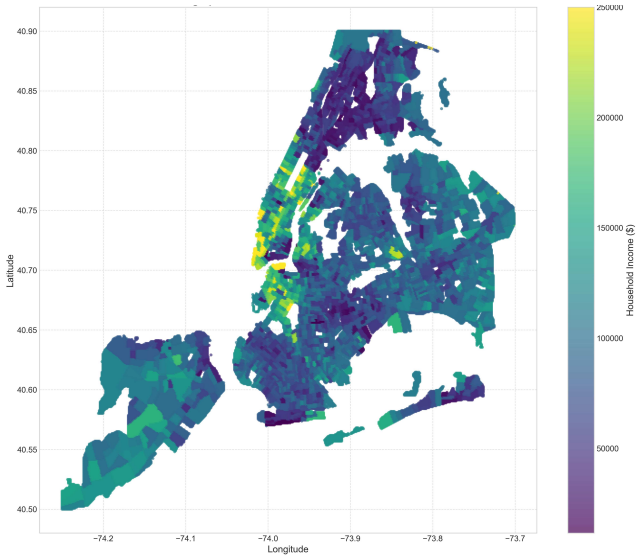

**Figure 2: Census-tract median household income across NYC. Colors denote income from low (purple) to high (yellow).**

## 5.3 Correlation Analysis

In this subsection, we summarize the analysis found from the Pearson correlation (shown in Figure 3).

**Race/ethnicity and income:** White population share is positively associated with household income ($r \approx +0.41$), whereas Black/African American population share is negatively associated ($r \approx -0.33$). Asian population share shows a modest positive association ($r \approx +0.10$); Native Hawaiian and Other Pacific Islander population share is weakly negative ($r \approx -0.06$); and Other single race population share and Multiracial population share are also negatively associated ($r \approx -0.41$ and $r \approx -0.16$, respectively).

**Race/ethnicity and CIAS:.** Non-library cultural accessibility scores (art gallery, theater, and museum CIAS) are highly intercorrelated ($r \approx 0.8-0.9$) and positively associated with population density, indicating higher values in denser tracts. By contrast, the library CIAS is more evenly distributed and only moderately correlated with the non-library CIAS ($r \approx 0.46-0.62$). But, its association with population density is the strongest among the CIAS components ($r \approx +0.58$), exceeding that of non-library CIAS.

Across racial/ethnic population shares, CIAS associations are small in magnitude. The Multiracial population share is positive across all four CIAS measures: gallery ($r \approx +0.08$), theater ($r \approx +0.07$), museum ($r \approx +0.14$), and library ($r \approx +0.17$). The African American population share is negative for the non-library CIAS: gallery ($r \approx -0.08$), theater ($r \approx -0.05$), museum ($r \approx -0.08$), but reverses sign for library ($r \approx +0.13$). The White population share shows the opposite pattern, positive with non-library CIAS: gallery ($r \approx +0.14$), theater ($r \approx +0.12$), museum ($r \approx +0.10$), and negative with library ($r \approx -0.25$). Other groups (e.g., Asian, American Indian/Alaska Native, Native Hawaiian and Other Pacific Islander, and Other single race) exhibit near-zero, non-significant correlations with the non-library CIAS (typically $|r| < 0.10$).

**CIAS and Household Income:** Pairwise correlations between cultural accessibility (CIAS) and median household income show modest positives for non-library infrastructures and a weak negative for libraries: art-gallery CIAS ($r \approx +0.29$), theater CIAS ($r \approx +0.22$), museum CIAS ($r \approx +0.22$), and library CIAS ($r \approx -0.15$). In short, tracts with higher non-library cultural accessibility tend to have higher median household incomes, whereas greater library accessibility is slightly associated with lower incomes.

**Population and income:** Total population per tract shows essentially no linear relationship with household income ($r \approx -0.02$), indicating that more populous tracts are not richer or poorer. By contrast, population density is slightly negatively correlated with median household income ($r \approx -0.23$). At the same time, density is positively aligned with cultural accessibility, as CIAS values increase with density across all infrastructure types, especially for libraries ($r \approx +0.58$). So denser tracts tend to have higher CIAS even though their median household incomes are lower.

**Depression and correlates:** Depression prevalence shows weak associations across features. It is negatively related to household income ($r \approx -0.20$) and slightly positively related to population density ($r \approx +0.19$). Correlations with cultural accessibility are small: art-gallery CIAS ($r \approx +0.07$), theater CIAS ($r \approx +0.07$), and museum CIAS ($r \approx +0.12$), with a comparatively stronger positive link for library CIAS ($r \approx +0.30$). Across racial/ethnic population shares, coefficients are generally small to moderate. Overall, depression aligns more with broad demographic context (e.g., density and population composition) than with cultural accessibility measures.

**Table 1: Model performance comparison. Best results are highlighted in bold.**

| Model | $R^2 \uparrow$ | RMSE $\downarrow$ | MAE $\downarrow$ | Parameter |
|---|---|---|---|---|
| Linear Regression | 0.4083 | 30118.41 | 22650.40 | 17 |
| **Random Forest** | **0.9127** | **11566.72** | **7771.45** | **4800** |
| Gradient Boosting | 0.6702 | 22484.62 | 16798.52 | 5040 |
| XGBoost | 0.7517 | 19511.74 | 14481.06 | 10800 |
| FCNN | 0.7611 | 19136.01 | 13580.19 | 3331 |

## 5.4 Model Performance and Interpretability

Our analysis reveals several significant findings regarding the prediction of household income. The Random Forest model emerged as the superior performer, achieving the highest $R^2$ score and the lowest errors (Table 1), significantly outperforming other models. Considering the "Parameter" column, Random Forest (4800) achieves the best accuracy–efficiency trade-off relative to more parameter-heavy alternatives, e.g., XGBoost (10800), while remaining easier to tune and generally more stable. These results support using tree ensembles as the primary predictive baseline.

From the SHAP analysis (Figure 4), the most influential feature is the "Other" race population share, whose higher values are associated with negative SHAP values (lower predicted income). The White population shows the next-largest contribution, with higher values contributing positively, whereas population density contributes negatively as density rises. The Black population share also has a substantial negative contribution at higher levels. Among CIAS, galleries and museums display the strongest positive contributions; the library and theater CIAS are small and negligible. Depression prevalence shows a weak negative contribution. Spatial coordinates (longitude and latitude) have moderate importance with mixed signs, capturing a central–peripheral gradient in the city. Overall, the model's attributions are dominated by demographic composition and density, with cultural accessibility playing a secondary, mostly positive role for only art galleries and museums (also visible in Figure 1b and 2).

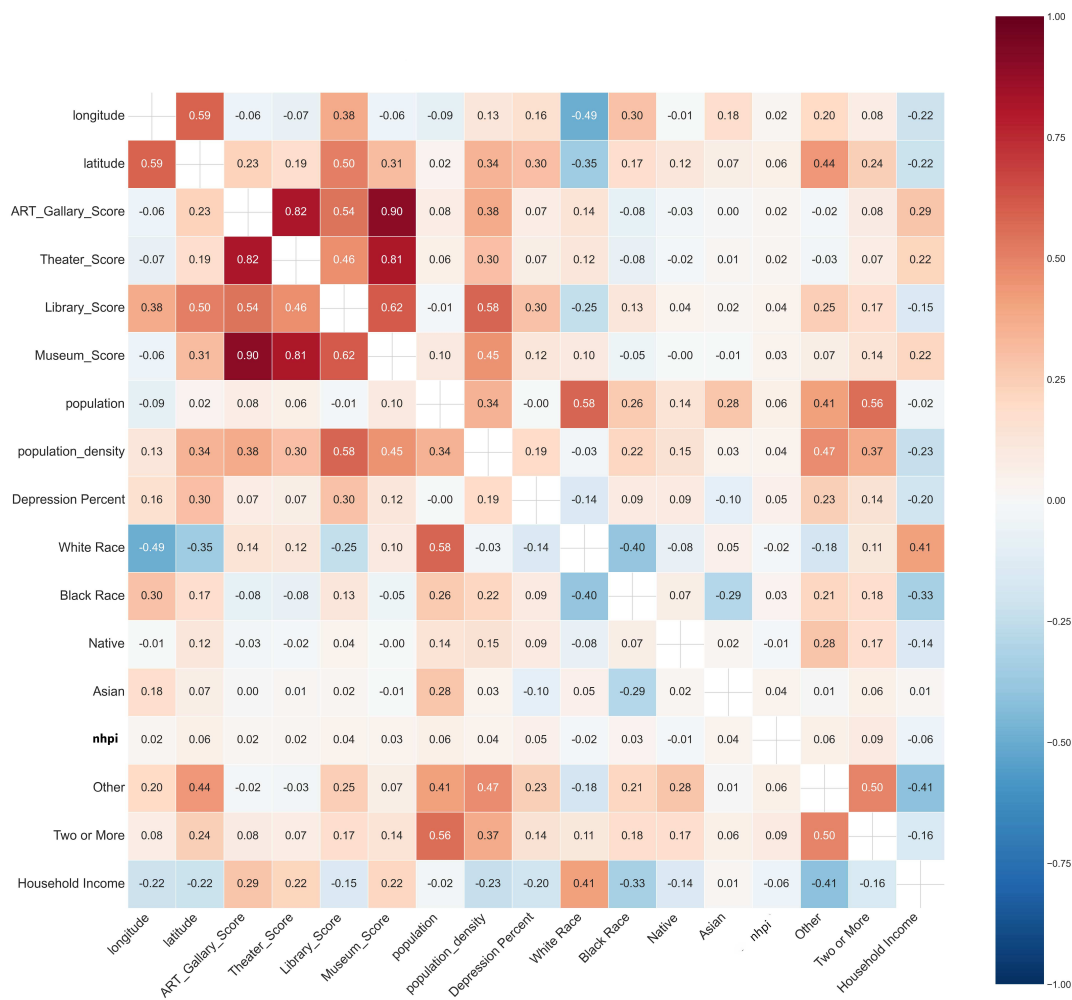

**Figure 3: Feature correlation matrix highlighting relationships between variables**

# 6 Discussion

## 6.1 Spatial Equity in Cultural Accessibility: Patterns and Drivers

Cultural access in New York City shows a pronounced center–edge distribution: resources are densest across much of Manhattan; taper to mixed coverage in Brooklyn (weakening toward the borough's outer edges); are heterogeneous and generally sparser in outer Queens; sit mostly mid-range in the Bronx; and remain sparse across most of Staten Island. This pattern mirrors the asset geography, too, where galleries, museums, and theaters cluster in the urban core (mostly Manhattan), while libraries provide the broadest coverage and extend farther into outlying neighborhoods. Some cultural infrastructures, such as galleries and museums, mostly cluster in higher-income regions. As income is unevenly distributed, this produces both place-level disparities (underserved vs. overserved regions) and unequal access across racial/ethnic groups.

*Place-level disparities:* Both the SCAS and the spatial distribution of cultural infrastructure indicate that overserved regions cluster across much of Manhattan and parts of Brooklyn. Underserved regions concentrate in outer Queens and in southern and eastern Brooklyn, and most of Staten Island. But, on the other hand, the Bronx is largely mid-range with pockets at both extremes. The pattern follows a core concentration with peripheral tapering.

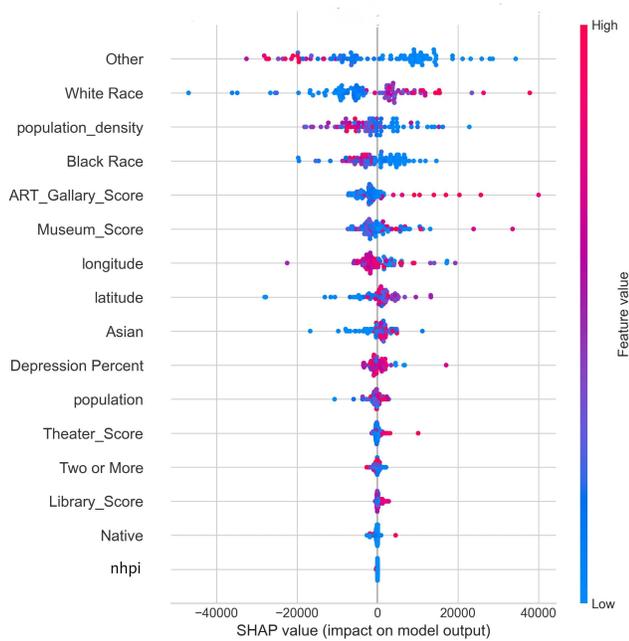

**Figure 4: SHAP summary plot showing feature importance for the Random Forest model. Features are ranked by their average impact on predictions across all samples. Positive SHAP values indicate higher income predictions; negative values indicate lower predictions.**

*Group-level disparities:* The same core–periphery gradient persists with group-specific nuances. White residents are most often overserved in Manhattan and central Brooklyn and increasingly underserved towards the outer boroughs. Black/African American residents exhibit a mixed pattern: some overserved tracts in Manhattan, but more underserved tracts in peripheral Queens, southern and eastern Brooklyn, and Staten Island. Asian residents are more likely to be overserved in Manhattan and central Brooklyn and underserved in outer Queens and Staten Island. Residents classified as "Other" are generally underserved toward the city's edges. American Indian/Alaska Native and Native Hawaiian/Other Pacific Islander populations show few clearly overserved tracts and more underserved tracts in the outer boroughs, where present. Multiracial residents are most often over-served in Manhattan and parts of central Brooklyn. Queens is mixed, with pockets of higher access nearer the core and increasing under-service toward the periphery. The Bronx is largely mid-range with scattered high-access tracts, and Staten Island is predominantly underserved.

*Why the disparities persist:* Correlations and model interpretability indicate a clear ordering of drivers consistent with observed access patterns. Socio-demographics mostly dominate: a higher White population share is associated with higher predicted income, a higher Black and "Other" shares with lower predicted income, and greater population density with lower income. Cultural proximity adds a secondary signal: gallery and museum accessibility correspond to higher predicted income, library and theater accessibility are small and contribute little. Two mechanisms help explain these patterns: (1) *economic sorting* — wealthier, more central tracts host or lie near clusters of galleries and museums, producing over-service at the core and under-service toward the edges; and (2) *urban-form moderation* — density aligns positively with proximity to cultural assets (especially libraries) yet negatively with income, yielding dense, lower-income neighborhoods that are relatively well served by libraries but under-served by art galleries and museums, while peripheral areas are distant from all assets. Descriptively, higher-income tracts tend to have higher gallery and museum CIAS and slightly lower library CIAS, whereas density is positively related to all CIAS components (strongest for libraries).

*Role of libraries:* Libraries are the notable counterpart. Their spatial distribution aligns more closely with population density than with income and is more uniform than that of art galleries, museums, or theaters. Consequently, dense, lower-income, racially diverse neighborhoods often exhibit comparatively higher library accessibility even when non-library accessibility is limited.

Together, these findings resolve our research questions and point to a coherent equity story: cultural access in NYC follows a core–periphery pattern with asset and group-specific heterogeneity. Variation in median household income is primarily associated with demographic composition and population density, while proximity to cultural assets adds a smaller, asset-specific association.

## 6.2 Implication

We recommend a place-based strategy that expands non-library cultural infrastructure beyond the Manhattan–Brooklyn core while using libraries as the everyday access backbone. Maps indicate a core–periphery gap for galleries, museums, and theaters, alongside broader library coverage in many dense neighborhoods. Accordingly, we propose to shift programs outward rather than adding more in the core, and to engage libraries as partners and hosts for outreach—not as substitutes for non-library venues. In dense neighborhoods just outside the core, where library access is strong but gallery/museum/theater access is weak, prioritize small, flexible sites and frequent programming (e.g., pop-up galleries in vacant storefronts, black-box performance rooms, and artist-in-residence studios co-located in schools, park fieldhouses, or community rooms). In the low-density periphery, where all forms of access score low, favor service models over new buildings (e.g., mobile cultural units, rotating "culture fairs" at transit nodes, and schedules aligned with evening and weekend transit). In outer Queens and Staten Island tracts with larger Asian populations, provide language-accessible curation, school partnerships, and programs aligned with cultural calendars. In the over-served core, tie a share of public funding to "export" commitments so anchor institutions' loan collections and send touring programs to non-core neighborhoods.

## 6.3 Limitations and Future Work

Cultural asset data come from open sources and may be incomplete or uneven across neighborhoods. We treat all venues within an asset type as equivalent, disregarding differences in size, capacity, hours, prices, and accessibility. Demographic and health indicators are available only at the census-tract scale, precluding within-tract heterogeneity, and sources might span different years, introducing potential temporal mismatch. Group-specific interpretations rely on tract composition, and the intra-tract spatial distribution of subgroups is unobserved. Our results describe associations, not causal effects. CIAS were not estimated with explicit spatial terms, so residual spatial dependence may remain, and the Modifiable Areal Unit Problem (MAUP) may affect relationships. Finally, distances in CIAS are computed as straight-line (Euclidean) measures. This overlooks the street and transit network, barriers, and time-of-day variation in travel. We will apply this framework to other cities of different sizes, transit systems, and cultural markets to see how well it works and adjust settings locally. We will use travel time on street and transit networks (by time of day) instead of straight-line distance, and check how results change when we vary key choices like decay rates, asset weights, grid size, and aggregation. We will add venue details (hours, pricing, capacity, accessibility, programming) and link access to actual use with attendance.

# 7 Conclusion

We introduced a novel, scalable, open-data framework to measure and explain spatial equity in cultural access. It maps cultural infrastructures, computes a Cultural Infrastructure Accessibility Score, and uses tree-ensemble models with SHAP to understand patterns in income, density, and tract composition. Accessibility follows a core–periphery structure, with non-library infrastructures clustering near urban cores, while libraries track density and reach farther across neighborhoods. Non-library access is modestly higher in higher-income tracts, and library access is slightly higher in denser, lower-income areas. These findings suggest adding non-library offerings in dense, lower-income neighborhoods beyond the core through satellites and pop-ups, using mobile and seasonal programs in low-density peripheries. Future work will add spatial models, network-time accessibility, richer venue attributes, and replication across multiple cities to strengthen guidance and transferability.